\documentclass[floatfix,reprint,amsmath,amssymb,aps,pra,
bibnotes]{revtex4-2}

\usepackage[hidelinks]{hyperref}
\usepackage[capitalize]{cleveref}
\usepackage{physics}
\usepackage{graphicx}
\usepackage{bm}
\usepackage{xspace}

\newcommand{\Usymm}{$U(1)$ symmetry\xspace}
\newcommand{\mean}[1]{\mathbb{E}[#1]}
\renewcommand{\d}{\mathrm{d}}
\newcommand{\diss}{\mathcal{D}}
\newcommand{\dt}{\frac{\mathrm{d}}{\mathrm{d}t}}
\newcommand{\phiab}{\phi_{AB}}
\newcommand{\gm}{\gamma_{-}}
\newcommand{\gp}{\gamma_{+}}
\newcommand{\rnoise}{x}
\newcommand{\inoise}{y}
\newcommand{\m}{\mathrm{m}}
\newcommand{\gainvdp}{\kappa_1}
\newcommand{\lossvdp}{\kappa_2}

\begin{document}
\date{April 10, 2026}

\title{Quantum limit cycles and synchronization from a measurement perspective}

\author{Tobias Nadolny}
\affiliation{Department of Physics, University of Basel, Klingelbergstrasse 82, 4056 Basel, Switzerland}
\author{Christoph Bruder}
\affiliation{Department of Physics, University of Basel, Klingelbergstrasse 82, 4056 Basel, Switzerland}

\begin{abstract}
Limit-cycle oscillators are the basic building blocks for synchronization; yet, the notion of a quantum limit cycle has remained unclear.
Here, we study quantum limit cycles and synchronization in the presence of continuous heterodyne measurement.
The resulting quantum trajectories, i.e., time evolutions of the quantum state conditioned on the measurement outcome, make the quantum limit cycles apparent.
We focus on the paradigmatic model of the quantum van der Pol oscillator and on two-level systems.
Our work provides insights into limit cycles in quantum systems, emphasizing their similarity to classical limit cycles subject to noise.
Additionally, we connect theoretical measures of quantum synchronization to quantities experimentally accessible via heterodyne detection.
\end{abstract}

\maketitle

\section{Introduction}
\label{s:synchronization}

Synchronization is the phenomenon whereby 
oscillating entities align their phases and entrain their frequencies even in the presence of frequency disorder and noise.
Synchronization occurs in various ways throughout nature, as well as in engineered and social systems.
Perhaps the most famous instance is synchronization of coupled clocks as originally discovered by Huygens~\citep{pikovsky_rosenblum_kurths_2001}; his experiment is still revisited centuries later~\citep{Bennett_2002,10.1063/5.0026335}.
Synchronization builds on self-sustained oscillators, 
in which gain and loss stabilize limit cycles.
Limit-cycle oscillators and synchronization have been extensively studied in classical nonlinear dynamics~\citep{pikovsky_rosenblum_kurths_2001,acebronKuramotoModelSimple2005,Strogatz_2019}.

The concepts of limit cycles and synchronization also extend to quantum systems.
Experimentally, quantum synchronization has been observed in cold atoms~\citep{Cox_2014,Weiner_2017,Natale_2025,PhysRevLett.125.013601}, nuclear spins~\citep{PhysRevA.105.062206}, trapped ions~\citep{zhang2023observing,Behrle_2023,Li_2025,Liu_2025}, and superconducting qubits~\citep{koppenhoferQuantumSynchronizationIBM2020,Tao_2025}.
Theoretical studies include the exploration of entanglement and other nonclassical features in synchronizing quantum systems~\citep{leeEntanglementTongueQuantum2014,rouletQuantumSynchronizationEntanglement2018,Zhu_2015,Witthaut2017,Lorenzo_2022,Fazio2013,PhysRevE.107.024204} as well as synchronization phenomena unique to quantum systems~\citep{lorchGenuineQuantumSignatures2016, lorchQuantumSynchronizationBlockade2017,amitaiQuantumEffectsAmplitude2018,Cooper2019,Mok_2020,Delmonte_2023,Chia2023,Nadolny_2023,Paul_2024,Kehrer_2024_2}.

Despite all of these results, the concept of a limit cycle in a quantum system has not been fully clarified~\citep{Talitha2017,Navarrete-Benlloch_2017,rouletSynchronizingSmallestPossible2018,Chia2020,ben_arosh_quantum_2021,parra-lopezSynchronizationTwolevelQuantum2020,Tan_2022,BucaJaksch2022,Setoyama_2024,Setoyama_2025,Dutta_2025,Zhao_2025,Chia_2025}.
This issue is often addressed on the level of the density operator and associated expectation values, which contain information about an ensemble of trajectories.
However, individual trajectories obtained in the presence of a continuous measurement can provide additional
insights~\citep{
Jacobs_2006,Wiseman_2009,Daley_2014}.
Quantum trajectories can be observed in various platforms including superconducting qubits~\citep{Gambetta_2008,Murch_2013},
optical cavities~\citep{Cox_2014,Weiner_2017,Yu_2024}, and
trapped ions~\citep{Bushev_2006}.
Some aspects of the role of measurement and trajectories in the context of quantum synchronization have been investigated previously.
For one, homodyne detection has been proposed to enhance synchronization with or without feedback~\citep{Kato_2021,Shen_2023}.
Moreover, an unraveling of the master equation into pure states has been employed to gain understanding of synchronization along individual quantum trajectories~\citep{Zhirov_2008,Weiss_2016,Eshaqi-Sani_2020} and to derive an effective phase equation~\citep{Setoyama_2024,Setoyama_2025}.
However, the full unraveling into pure states that is considered in previous studies is usually not experimentally feasible, as it requires perfect monitoring of all dissipative processes.

In this paper, we theoretically study the unraveling of a single jump operator describing the loss of excitations.
We consider heterodyne detection because it continuously measures the oscillator's amplitude, connecting it with classical trajectories.
This unraveling allows for the real-time observation of a single quantum limit cycle or several synchronizing oscillators.
We demonstrate that continuous heterodyne detection results in quantum trajectories that make quantum limit cycles and their synchronization apparent:
The trajectories feature the signature dynamics of classical limit cycles subject to noise that are not directly visible in the time evolution of the ensemble-averaged density operator. 
This contributes to the understanding of the nature of quantum limit cycles.
Furthermore, continuous monitoring results in experimentally accessible quantities that approximate established theoretical measures of quantum synchronization.
Indeed, heterodyne detection has already been applied in experiments to measure synchronization in atomic ensembles~\citep{Cox_2014,Weiner_2017,Natale_2025}.

We begin by reviewing the synchronization of classical oscillators subject to fluctuations in \cref{s:classical_synchronization}.
In \cref{s:quvdp}, we will analyze the quantum van der Pol oscillator.
We first show how the heterodyne measurement makes the limit cycle apparent in \cref{s:qvdp_qlc}.
In \cref{s:qvdp_2}, the synchronization of two such oscillators is analyzed.
An analogous analysis of synchronization among quantum two-level systems, i.e., spins-$1/2$, is presented in \cref{s:quspins}, starting with a single spin and progressing to two coupled spins.
We conclude in \cref{s:conclusions}.

\section{Classical Synchronization}
\label{s:classical_synchronization}

We start by introducing the basic concepts of synchronization in classical systems.
In \cref{s:limitcycleoscillator}, we show how a single limit-cycle oscillator can be described in terms of its phase.
Phase locking and frequency entrainment of two coupled oscillators are discussed in \cref{s:sync_two_coupled}.

\subsection{Limit-cycle oscillator}
\label{s:limitcycleoscillator}

A paradigmatic model of a limit-cycle oscillator is the van der Pol (vdP) oscillator.
In the regime of weak nonlinearity, the oscillator's 
amplitude $\alpha$ follows the equation of motion~\citep{pikovsky_rosenblum_kurths_2001}
\begin{equation}
    \dot \alpha = -i \omega \alpha + \gainvdp \alpha / 2 - \lossvdp \abs{\alpha}^2 \alpha
    \, ,
    \label{eq:vdP}
\end{equation}
with linear gain at rate $\gainvdp>0$ and nonlinear loss at rate $\lossvdp>0$.
The limit-cycle trajectory can be obtained by defining the amplitude $r = \abs{\alpha}$ and phase $\phi=-\arg[\alpha]$, 
so that
\begin{align}
        \dot r &= (\gainvdp/2 - \lossvdp r^2) r
        \, ,
        \quad
        \dot \phi = \omega \, .
        \label{eq:vdp_phase} 
\end{align}
The radius is attracted to the stable fixed point at $r_0 = \sqrt{\gainvdp/2\lossvdp}$, and the phase increases linearly in time at rate $\omega$.
Consequently, the vdP oscillator exhibits an 
attractive limit cycle.
\begin{figure}
    \centering
    \includegraphics[width=2in]{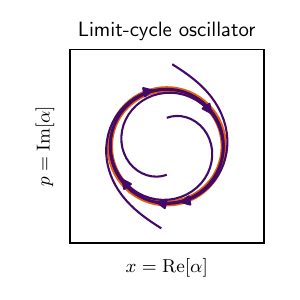}
    \caption{
    Four trajectories of a van der Pol oscillator, \cref{eq:vdP}, with $\kappa_1 = \kappa_2 = \omega / 2$.
    The red circle shows the limit cycle to which all trajectories converge.}
    \label{fig:HO_vs_LC}
\end{figure}
\Cref{fig:HO_vs_LC} shows four trajectories with different initial conditions approaching this limit cycle.

\subsubsection{Influence of fluctuations}
\label{s:LC_noise}
In the presence of fluctuations, the time evolution of the vdP oscillator can be modeled by the Langevin equation
\begin{equation}
    \dot \alpha = -i \omega \alpha + \gainvdp \alpha / 2 - \lossvdp \abs{\alpha}^2 \alpha
    + \sigma \xi_\rnoise (t)
     + i \sigma \xi_\inoise (t)
    \, .
    \label{eq:vdP_noise}
\end{equation}
The terms $\xi_\rnoise(t)$ and $\xi_\inoise(t)$ are stationary Gaussian white-noise processes that induce independent fluctuations in the real and imaginary parts of the amplitude $\alpha$.
The noise processes have zero mean, $\mean{\xi_\rnoise(t)} = \mean{\xi_\inoise(t)} = 0$, and their variances are $\mean{\xi_\rnoise(t)\xi_\rnoise(t')} = \mean{\xi_\inoise(t)\xi_\inoise(t')} = \delta(t-t')$.
The strength of the noise is $\sigma^2$.
For simplicity, we only consider Gaussian white noise here; for the influence of other types of noise on synchronization, see for example Ref.~\cite{Bag_2007}.

\begin{figure}
    \centering
    \includegraphics[width=3.4in]{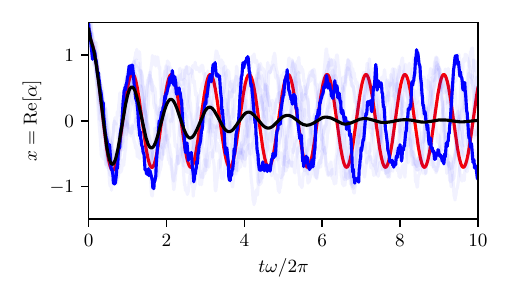}
    \caption{Limit-cycle oscillations in the presence of noise.
    The blue lines show $x=\Re[\alpha]$ for ten different trajectories, all starting with the same initial condition.
    One of them is highlighted in a darker blue.
    The black line is obtained by averaging over 16\,000 trajectories.
    The red line shows a trajectory in the absence of noise.
    Parameters: $\kappa_1 = \kappa_2 = \omega / 2 = 10\sigma^2$.
    }
    \label{fig:vdp_classical_trajectories}
\end{figure}

A few trajectories obtained by integrating \cref{eq:vdP_noise} are displayed in \cref{fig:vdp_classical_trajectories} (blue lines).
They approximately follow the noiseless evolution (red line) with additional fluctuations.
Since the information about the initial phase is lost due to the fluctuations,
the average of many trajectories (black line) approaches zero in the long-time limit.

Equation~\eqref{eq:vdP_noise} corresponds to the Fokker-Planck equation~\citep{Risken_1996}
\begin{equation}
    \begin{split}
    &\partial_t P (x,p) = - \bm{\nabla} (\bm{\mu} P(x,p) ) + \sigma^2 \bm{\nabla}^2 P(x,p)  \, ,
    \\
    \bm{\nabla} = &
    \begin{pmatrix}
        \partial_x \\ \partial_p
    \end{pmatrix}
    ,
    \, \, 
    \bm{\mu} = 
    \begin{pmatrix}
         \omega p + \kappa_1 x / 2 - \kappa_2 (x^2+p^2) x \\ 
        -\omega x + \kappa_1 p / 2 - \kappa_2 (x^2+p^2) p
    \end{pmatrix}
    \, ,
    \end{split}
    \label{eq:FP_vdP} 
\end{equation}
where we introduced the probability distribution $P(x,p)$ for the vdP oscillator to be in the state $\alpha = x + ip$ at time $t$.
The deterministic part of the time evolution enters via the term proportional to $\bm\mu$, while the effect of the fluctuations is described by the term proportional to $\sigma^2$.

\begin{figure*}
    \centering
    \includegraphics[width=5.8in]{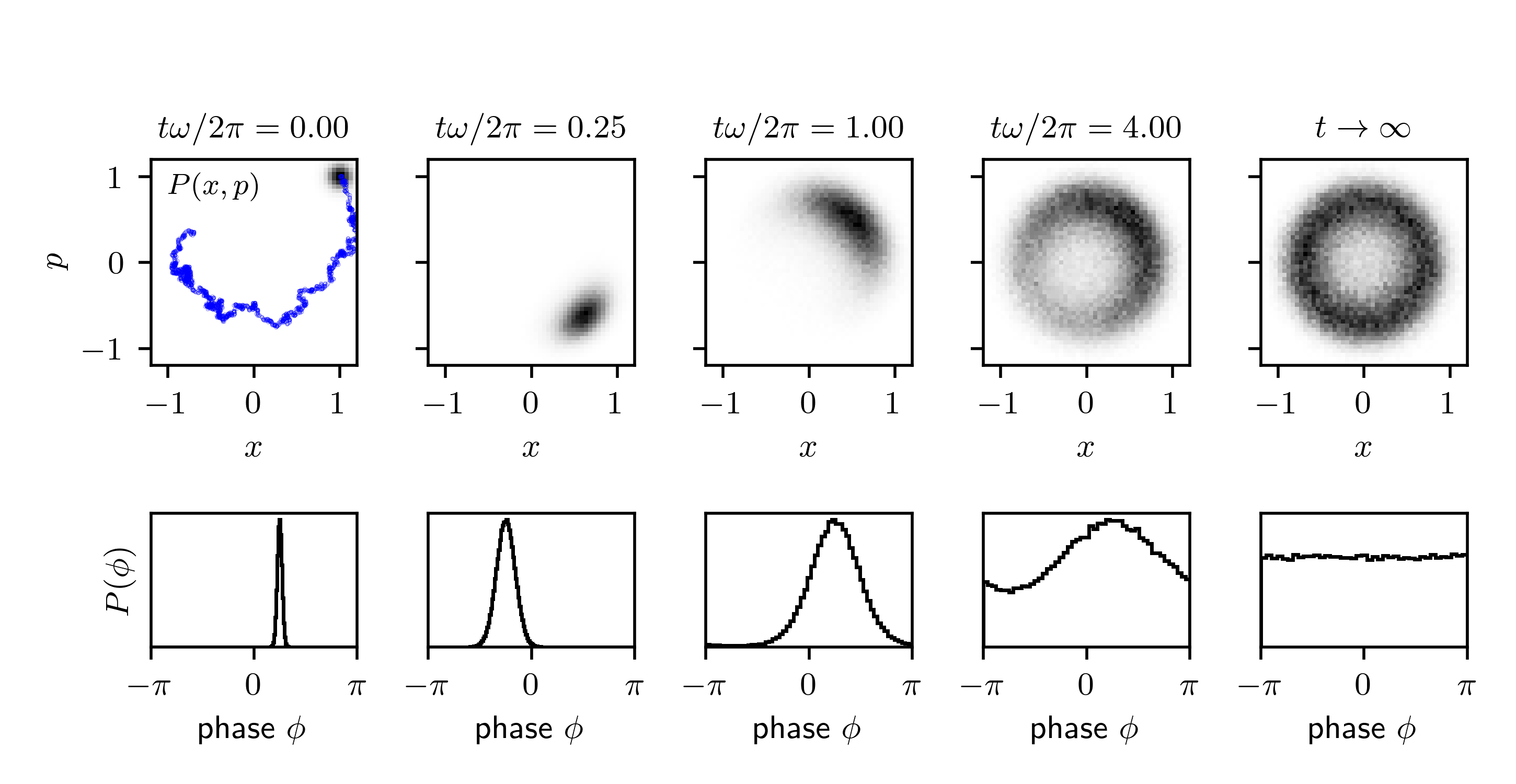}
    \caption{
    Time evolution of the classical vdP oscillator shown by the probability distributions $P(x,p)$ (top row) and $P(\phi)$ (bottom row).
    In the top row, the grayscale indicates the value of $P(x,p)$.
    The distributions are obtained by sampling 16\,000 trajectories and counting the occurrences per $(x,p)$ or per $\phi$.
    The blue line in the top left panel shows one example of a trajectory.
    Parameters: $\kappa_1 = \kappa_2 = \omega / 2 = 10 \sigma^2$.
    }
    \label{fig:vdP_noise}
\end{figure*}

For an approximate solution of \cref{eq:FP_vdP}, we integrate the corresponding Langevin equation, \cref{eq:vdP_noise},
for various initial conditions and realizations of the noise.
Given sufficient realizations, 
$P(x,p)$ is well approximated by the relative number of occurrences of the values $\alpha = x+ip$ at each time counted over all trajectories.
The time evolution of the probability distribution is shown in the upper panels of \cref{fig:vdP_noise}.
One recognizes the oscillation at frequency $\omega$ and the attraction to the limit cycle since the probability distribution rotates and moves towards the limit-cycle amplitude $r_0$.
Furthermore, the probability distribution spreads along the phase direction, indicating that the initial phase diffuses over time.
The initial distribution approaches a unique stationary distribution in the long-time limit, which is ring shaped and phase symmetric, i.e., invariant under rotations around the origin; see top-right panel of \cref{fig:vdP_noise}.

The time evolution of the phase distribution $P(\phi)$ obtained by integrating $P(x,p)$ over the radius is shown in the bottom row of \cref{fig:vdP_noise}.
Here, phase diffusion becomes visible as a broadening of $P(\phi)$, which approaches a flat curve in the long-time limit.

Another quantity that can be used to analyze the limit cycle's properties is the spectrum, which is defined as the Fourier transform of the steady-state two-time correlations
\begin{equation}
    g(\tau) \equiv \lim_{t\rightarrow\infty} \mathbb{E}[\alpha^*(t+\tau)\alpha(t)] \, .
    \label{eq:LC_correlation1}
\end{equation}
The expectation value $\mathbb{E}[\cdot]$ denotes an average over all possible noise realizations.
When the amplitude is large, one obtains~\citep[Section~11.4]{Scully_1997}
\begin{equation}
    g(\tau) = r_0^2 \exp(i\omega \tau - \frac{\sigma^2}{2r_0^2} \abs{\tau})
    \, . 
    \label{eq:LC_correlation3}
\end{equation}
The spectrum, i.e., the Fourier transform of $g(\tau)$, is a Lorentzian distribution centered at frequency $\omega$ with width ${\sigma^2}/{r_0^2}$.
The position of the peak corresponds to the frequency of the limit cycle, while the spectral linewidth is determined by the fluctuations.

\begin{figure*}
    \centering
    \includegraphics[width=2.8in]{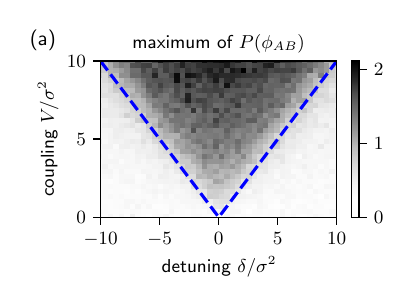}%
    \includegraphics[width=3in]{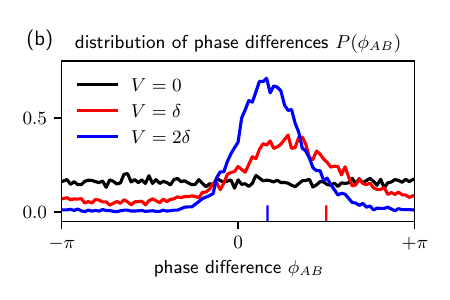}
    \caption{Phase locking of classical vdP oscillators.
    (a) Arnold tongue.
    The blue dashed line indicates the synchronization transition $V = \abs{\delta}$.
    The grayscale shows the maximum of $P(\phiab)$, a measure for synchronization in the presence of noise.
     For each value of the detuning and the coupling strength, the distribution is obtained from a trajectory of duration $10^3/\sigma^{2}$ with $10^3$ uniformly spaced samples.
    (b) Distribution $P(\phiab)$ of the phase difference in the long-time limit in the presence of noise and detuning, $\sigma^2=\delta$.
    The blue and red ticks indicate the value of the phase $\phiab^\mathrm{f}=\arcsin{\delta/V}$ in the absence of noise.
     For each value of the coupling strength, the distribution is obtained from a trajectory of duration $10^4/\delta$ with $10^4$ uniformly spaced samples.
    }
    \label{fig:classical_phase_locking}
\end{figure*}

\subsection{Two coupled oscillators}
\label{s:sync_two_coupled}
In this section, we show that two limit-cycle oscillators that are coupled strongly enough can entrain their frequencies and lock their phases despite the presence of frequency detuning or noise.

\subsubsection{Frequency detuning}
\label{ss:adler_detuning}

We start the analysis assuming zero fluctuations.
The evolution of two detuned and dissipatively coupled vdP oscillators is~\citep[Section 8.2]{pikovsky_rosenblum_kurths_2001}
\begin{subequations}
\begin{align}
  \dot \alpha &= -i \delta \alpha/2 + \gainvdp \alpha / 2 - \lossvdp \abs{\alpha}^2 \alpha
  + V (\beta- \alpha) / 2
  \\
  \dot \beta &= +i \delta  \beta/2 + \gainvdp \beta / 2 - \lossvdp \abs{\beta}^2 \beta
  + V (\alpha- \beta) / 2
\end{align}
\label{eq:2vdp_alpha}%
\end{subequations}
with frequency detuning $\delta$ and coupling strength $V \geq 0$.
In writing Eqs.~\eqref{eq:2vdp_alpha}, we have implicitly moved to a frame rotating at the average frequency of the two oscillators. 
In that frame, the frequencies of the oscillators are $\pm \delta/2$.

We define amplitudes and phases of the oscillators via $\alpha = r_{A} \exp(-i\phi_{A})$ and $\beta = r_{B} \exp(-i\phi_{B})$.
The phase difference $\phiab=\phi_A - \phi_B$ plays a central role; its time evolution is  
\begin{equation}
    \dot{\phi}_{AB} = \delta - \frac{V}{2} \left(\frac{r_A}{r_B} + \frac{r_B}{r_A} \right) \sin(\phiab)
    \, .
    \label{eq:adler_pre}
\end{equation}
In the context of synchronization, the rates $\kappa_{1}$ and $\kappa_{2}$, which stabilize the limit cycle, are considered large compared to the coupling strength $V$.
In this case, we can approximate the radius as constant with $r_{A,B} = r_0 =\sqrt{\kappa_1/2\kappa_2}$.
\Cref{eq:adler_pre} consequently becomes the Adler equation~\citep{adler,pikovsky_rosenblum_kurths_2001}
\begin{equation}
    \dot{\phi}_{AB} = \delta - V\sin(\phiab)    
    \label{eq:adler}
\end{equation}
that captures the essential dynamics of two coupled limit-cycle oscillators.
It describes a competition between the detuning, which causes the phase difference to grow, and the coupling, which brings the phase difference closer to zero. 

When the detuning is smaller than the coupling, ${\abs{\delta} <V}$, the Adler equation exhibits a stable fixed point
 \begin{equation}
     \phiab^\mathrm{f} = \arcsin(\delta / V) \, .
     \label{eq:classical_phase_locking}
 \end{equation}
In this case, the two oscillators synchronize: They oscillate at the same frequency and their phase difference is locked to a constant value.
The region in which synchronization occurs is called the Arnold tongue; the boundary between the synchronized and unsynchronized regimes, $\abs{\delta} = V$, is shown in \cref{fig:classical_phase_locking}(a) by the blue dashed line.

\begin{figure*}
    \centering
    \includegraphics[width=2.8in]{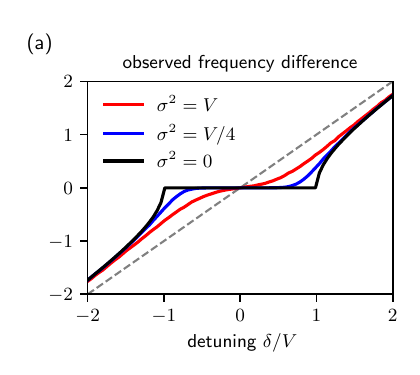}%
    \includegraphics[width=3in]{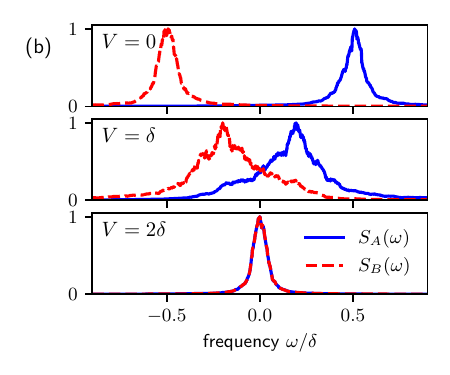}
    \caption{
    Frequency entrainment of classical vdP oscillators.
    (a)
    Observed frequency difference as a function of detuning.
    The dashed line shows the identity line for reference.
    The black line shows the noiseless case ($\sqrt{\delta^2/V^2-1}$).
    The blue and red lines show the observed frequency for two different noise strengths.
    Although difficult to see, the observed frequency is not exactly zero for any $\sigma^2 > 0$ and $\abs{\delta}>0$.
    (b)
    Spectra of the two phase oscillators for $\sigma^2 = \delta/5$, as defined in \cref{eq:spectra_classical} normalized to their maximum value.
    Each spectrum is averaged in bins of width $\omega/\delta = 0.05$.
     For each value of the coupling strength, the spectra are obtained from a trajectory of duration $10^4/\delta$ with $10^4$ uniformly spaced samples.
    }
    \label{fig:classical_frequency_entrainment}
\end{figure*}

When the detuning is larger than the coupling strength, $\abs{\delta}>V$, the two oscillators are unsynchronized.
Solving the Adler equation yields an average observed frequency difference $\sqrt{\delta^2 - V^2}$~\citep{pikovsky_rosenblum_kurths_2001},
which is shown by the black line in \cref{fig:classical_frequency_entrainment}(a).

\subsubsection{Influence of fluctuations}
\label{ss:adler_noise}
In the presence of noise, phase locking and frequency entrainment are approximate rather than exact, as we now show.
The equations for the phases derived from Eqs.~\eqref{eq:2vdp_alpha} with added noise at strength $\sigma^2/2$ (setting again $r_{A,B} = r_0$) are
\begin{subequations}
\begin{align}
    \dot \phi_A = +\frac{\delta}{2} + \frac{V}{2} \sin(\phi_B - \phi_A) + \frac{\sigma}{\sqrt{2}} \xi_A
    \, ,
    \\
    \dot \phi_B = -\frac{\delta}{2} + \frac{V}{2} \sin(\phi_A - \phi_B) + \frac{\sigma}{\sqrt{2}} \xi_B
    \, .
    \end{align}
    \label{eq:2vdp_phases_noise}%
\end{subequations}
The two independent Gaussian white-noise processes $\xi_A$ and $\xi_B$ that cause fluctuations in the phases have zero mean and variance $\mean{\xi_{A,B}(t)\xi_{A,B}(t')} = \delta(t-t')$.

We can quantify the amount of phase locking using the distribution $P(\phiab$) of the phase difference in the long-time limit.
Numerically, this distribution can be obtained by integrating Eqs.~\eqref{eq:2vdp_phases_noise} in the presence of noise for various initial conditions and counting how often $\phiab$ assumes a certain value in the long-time limit.
The results are shown in \cref{fig:classical_phase_locking}(b) for different values of the coupling.
When the coupling is absent, the phase distribution is flat, showing that all phase differences are equally likely.
A nonzero coupling induces a phase preference indicated by a peak in the phase distribution.
The phase is more likely to take values close to $\phiab^\mathrm{f} = \arcsin(\delta/V)$, the stable fixed point obtained in the noise-free analysis; see \cref{eq:classical_phase_locking}.

\Cref{fig:classical_phase_locking}(a) shows the maximum value of $P(\phiab)$ as a function of detuning and coupling strength.
The maximum grows with increasing coupling and decreasing detuning, and displays a crossover at the synchronization transition $V=\abs{\delta}$ obtained in the noiseless case (blue dashed line).

To further analyze the approximate frequency entrainment in the presence of noise, we consider the spectrum of the oscillators.
The steady-state two-time correlations, introduced in Eq.~\eqref{eq:LC_correlation1}, are 
\begin{equation}
    g_a(\tau) = \lim_{t\rightarrow\infty} \mathbb{E}[\exp[i\phi_a(t+\tau)-i\phi_a(t)]] \, ,
\end{equation}
where $a \in \{A,B\}$.
The spectra, which depend on frequency $\omega$, are obtained via Fourier transformation,
\begin{equation}
    S_a(\omega) = \int_{-\infty}^\infty \mathrm{d}\tau e^{-i\omega\tau} g_a(\tau) \, .
    \label{eq:spectra_classical}
\end{equation}

We numerically calculate the spectra by averaging over multiple trajectories obtained from integrating Eqs.~\eqref{eq:2vdp_phases_noise}.
The spectra are shown in \cref{fig:classical_frequency_entrainment}(b) for different values of the coupling strength. 
They are characterized by a peak whose position indicates the typical frequency and whose width is determined by the phase diffusion.
For zero coupling, the spectra both peak at the natural frequencies $\pm \delta/2$.
With increasing coupling, they peak at frequencies closer to zero until they nearly overlap.
This behavior is consistent with that of the average frequency difference shown in  \cref{fig:classical_frequency_entrainment}(a).

The frequency spectra and the phase distribution $P(\phiab)$ facilitate the quantification of frequency entrainment and phase locking even when noise prohibits exact synchronization.
In \cref{s:quvdp} and \cref{s:quspins}, we will encounter analogous measures of quantum synchronization.

\section{Quantum van der Pol oscillator}
\label{s:quvdp}

We begin the analysis of quantum synchronization by presenting a quantum analog of the classical van der Pol (vdP) oscillator.
The quantum vdP oscillator is described by ladder operators $a^\dag$ and $a$ that add or remove an excitation of a bosonic mode and fulfill the commutation relation $[a,a^\dag]=1$.
Gain and loss of the oscillator are introduced through the coupling to an (unspecified) environment in the framework of open quantum systems~\citep{Breuer}.
The Lindblad master equation of the quantum vdP oscillator~\citep{lee_QuantumSynchronizationQuantum_2013,walterQuantumSynchronizationDriven2014} (also referred to as Stuart-Landau oscillator~\citep{Chia2020,Chia_2025} or quantum Rayleigh-van der Pol oscillator~\citep{ben_arosh_quantum_2021}) is
\begin{equation}
    \label{eq:qvdp}
    \dt \rho =
    -i[\omega a^\dag a, \rho ]
    + \kappa_1 \diss[a^\dag]\rho + \kappa_2 \diss[a^2]\rho
    + \kappa \diss[a]\rho \, .
\end{equation}
The Lindblad dissipator is $\mathcal{D}[o]\rho = o\rho o^\dag -(o^\dag o \rho + \rho o^\dag o)/2$.
The oscillator's frequency is $\omega$, and $\kappa_1$, $\kappa_2$, and $\kappa$ are the rates of linear gain, nonlinear two-excitation loss, and linear loss, respectively.
The master equation is invariant under a phase shift, $a\rightarrow a \exp(i\phi_0)$, which corresponds to a \Usymm.

The connection between the quantum and classical vdP oscillators is evident in the time evolution of the oscillator's amplitude $\expval{a} \equiv \Tr[a\rho]$,
\begin{equation}
    \dt \expval{a} = 
    -i \omega \expval{a} + \frac{\kappa_1 - \kappa}{2} \expval{a} - \lossvdp \expval*{a^\dag a^2} \, .
    \label{eq:mf_qvdp1}
\end{equation}
When the state is initialized as a coherent state $\ket{\alpha}$ and assuming that it remains a coherent state, one can approximate the last term in \cref{eq:mf_qvdp1}:
$\expval*{a^\dag a^2} \approx \abs{\expval{a}}^2\expval{a}$.
The assumption is valid for short time scales compared to the strength of decoherence, where fluctuations play a negligible role.
Within this approximation, one obtains the equation of motion for a vdP limit-cycle oscillator presented in \cref{eq:vdP} (there, we have set $\kappa=0$).
In general, however, this approximation does not hold, and therefore, in the following, we analyze the full master equation.

\subsection{Quantum limit cycles}
\label{s:qvdp_qlc}

\begin{figure*}
    \centering
    \includegraphics[width=5.8in]{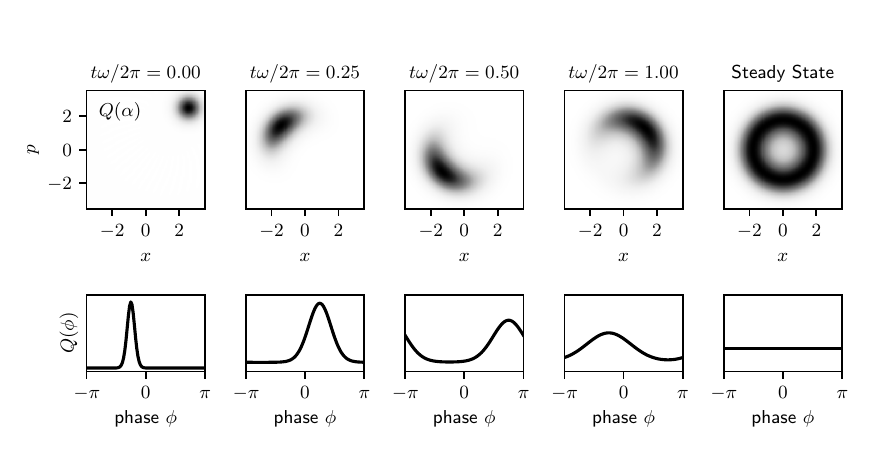}
    \caption{Time evolution of the quantum vdP oscillator shown by the Husimi-Q distribution $Q(\alpha=x+ip)$ (top row) and the phase distribution $Q(\phi)$ (bottom row).
    Parameters: $\kappa_1 = \omega = 4\kappa$, $\kappa_2 = \kappa/2$.
    }
    \label{fig:qvdp_time_evolution}
\end{figure*}
A numerical integration of the master equation~\eqref{eq:qvdp} yields the time evolution of the density operator.
To display the time evolution in phase space, the density operator is projected onto coherent states, which gives the Husimi-Q distribution~\citep{carmichael_springer_1}
\begin{equation}
    Q(\alpha) = \frac{1}{\pi}\expval{ \rho }{ \alpha } \, .
    \label{eq:qvdP_Q}
\end{equation}
We choose the Husimi-Q distribution because it is analogous to the classical phase-space distribution and allows us to treat the description of limit cycles and phase locking of classical and quantum vdP oscillators as well as of spins (as we will see in \cref{s:quspins}) in a parallel way.

The time evolution of the quantum vdP oscillator in terms of the Husimi-Q distribution is shown in \cref{fig:qvdp_time_evolution}.
The dynamics can be understood as three simultaneous processes.
First, the initial coherent state oscillates at frequency $\omega$, which can be seen as a rotation in time around the origin.
Second, as a result of gain and nonlinear loss, the amplitude of the oscillation is attracted to the limit cycle, and the radial position of the state approaches the value $r_0 = \sqrt{\kappa_1/2\kappa_2}$.
Third, the state diffuses in phase due to the coupling to the environment.
The attraction towards the limit cycle and the phase diffusion together result in a stationary ring-shaped state in the long-time limit, see the last panel of \cref{fig:qvdp_time_evolution}.
The steady state is phase symmetric, respecting the \Usymm of the master equation.

Comparing \cref{fig:qvdp_time_evolution} to \cref{fig:vdP_noise}, we find that the dynamics of $Q(\alpha)$ are qualitatively the same as those of the probability distribution $P(x,p)$.
Similar to how $P(x,p)$ describes an ensemble of trajectories, we will later see how $\rho$ -- and consequently $Q(\alpha)$ -- describes an ensemble of quantum trajectories.

The phase distribution $Q(\phi)$ displayed in the bottom row in \cref{fig:qvdp_time_evolution} is obtained by integrating out the radial degree of freedom,
$
    Q(\phi) =
    \int_0^{\infty} \d r \, r \, 
    Q(\alpha = r e^{-i\phi})
    \, .
    \label{eq:qvdP_Q_phasedistribution}
$
It is the probability of the quantum vdP oscillator to assume a certain phase $\phi$.
The phase distribution of the initial state is sharply peaked.
Over time, it shifts due to the oscillation and flattens due to the phase diffusion approaching a flat distribution in the steady state.
The phase distribution $P(\phi)$ of the classical vdP oscillator qualitatively showed the same behavior; see \cref{fig:vdP_noise}.

\subsubsection{Quantum limit cycles under heterodyne detection}

\begin{figure*}
    \centering
    \includegraphics[width=5.8in]{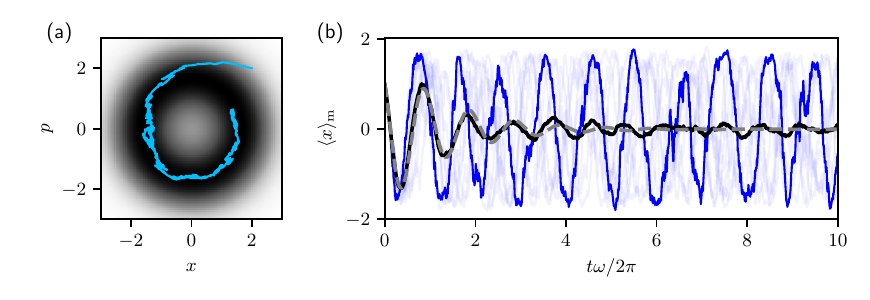}%
    \caption{Quantum vdP oscillator: limit cycle under measurement.
    (a)~The heatmap shows the steady-state Husimi-Q distribution from \cref{fig:qvdp_time_evolution}, top-right panel.
    The blue line shows the time evolution of $\expval{x}_\m$ and $\expval{p}_\m$ of one quantum trajectory.
    (b)~The light blue lines show the expectation value $\expval{x}_\m$ of ten different quantum trajectories.
    The dark blue line highlights one of them.
    The black line shows the average of $100$ such realizations.
    It overlaps well with the gray dashed line, which shows $\expval{x}$ obtained from the solution of the master equation without measurement.
    Same parameters as in \cref{fig:qvdp_time_evolution}.
    }
    \label{fig:qvdp_LC_measurement}
\end{figure*}

While the steady state shown in the top-right panel of \cref{fig:qvdp_time_evolution} resembles the classical limit cycle in its circular shape, it is static and appears not to feature the dynamical oscillations of a classical limit cycle.
The reason is that the density operator $\rho$ describes the probability distribution of an ensemble of quantum trajectories.
The steady-state probability distribution of an ensemble of classical trajectories in the presence of noise is also time-independent; see \cref{fig:vdP_noise}.
We now show that individual quantum trajectories obtained through heterodyne detection display dynamical limit-cycle oscillations in the long-time limit.

In heterodyne detection, the leaking excitations are mixed with a local oscillator with frequency $\omega_\m$ that is large compared to the system's frequency~\citep{Wiseman_2009}.
By monitoring the mixed signal, two quadratures are effectively measured simultaneously so that one can extract information about the system's amplitude.

The stochastic master equation~\citep{Wiseman_2009} for the quantum vdP oscillator under heterodyne detection is 
\begin{equation}
    \begin{split}
    \label{eq:1qvdp_meas}
    \dt &\rho_\m =
    -i[ \omega a^\dag a, \rho_\m ]
    + \kappa_1 \diss[a^\dag]\rho_\m +
    \kappa_2 \diss[a^2]\rho_\m
    + \\
    & + 
    \kappa \diss[a] \rho_\m +
    \frac{\mathrm{d}W}{\d t}
    \sqrt{\kappa} 
    \left[
    e^{i\omega_\m t}(a - \Tr[a\rho_\m]) \rho_\m + \mathrm{H.c.}
    \right]
    \, .
    \end{split}
\end{equation}
The density operator $\rho_\m$ denotes the conditional state of the system under measurement.
The measurement backaction that results in fluctuations in the state of the system is described by the term proportional to the Gaussian noise process $\d W/\d t$ with zero mean, $\mean{\mathrm{d}W}=0$, and variance $\mean{\mathrm{d}W^2}=\mathrm{d}t$.

The system's time evolution conditioned on a measurement outcome is called a quantum trajectory.
Each quantum trajectory is different because it depends on the measurement backaction, i.e., the random fluctuations induced in the system due to the measurement.
Averaging over the ensemble of possible quantum trajectories reproduces the unconditional evolution of the density operator governed by \cref{eq:qvdp}.
The density operator may thus be viewed as describing the ensemble of all quantum trajectories averaged over all possible measurement outcomes.

We simulate the time evolution governed by \cref{eq:1qvdp_meas} to obtain individual quantum trajectories; they are shown in \cref{fig:qvdp_LC_measurement}.
Up to fluctuations, the amplitude of the quantum vdP oscillator in each quantum trajectory displays a limit-cycle behavior: It is attracted to the limit cycle amplitude [see \cref{fig:qvdp_LC_measurement}(a)] and displays persistent oscillations [see \cref{fig:qvdp_LC_measurement}(b)].
While the steady-state density operator obeys the \Usymm of the master equation, each trajectory breaks it since it displays a distinct phase at any point in time.
Due to phase diffusion, the average of several trajectories decays to zero; see the black line in \cref{fig:qvdp_LC_measurement}(b).
It agrees well with the density matrix evolution shown by the gray dashed line in \cref{fig:qvdp_LC_measurement}(b).

The quantum trajectories obtained under heterodyne detection show qualitatively the same behavior as the trajectories of the classical vdP oscillator; see \cref{fig:vdp_classical_trajectories}.
Consequently, many aspects of quantum vdP oscillators can be qualitatively understood in terms of classical noisy oscillators.
One example is synchronization, which will be explored in the following section.

\subsection{Two coupled oscillators}
\label{s:qvdp_2}

Having introduced the quantum vdP oscillator, we now examine how two such oscillators can synchronize when coupled.
The two quantum vdP oscillators are described by the master equation
\begin{equation}
    \label{eq:2qvdp}
    \begin{split}
    \dt \rho =
    &-i\frac{\delta}{2}[ a^\dag a - b^\dag b, \rho ]
    + V \diss[a - b]\rho 
    \\
    &+\kappa_1 (\diss[a^\dag] +\diss[b^\dag])\rho
    + \kappa (\diss[a] +\diss[b])\rho 
    \\
    &+
    \kappa_2 (\diss[a^2]+\diss[b^2])\rho
     \, .
    \end{split}
\end{equation}
The density operator $\rho$ describes the joint state of both oscillators represented by the operators $a^{(\dag)}$ and $b^{(\dag)}$.
The master equation describes the system in the frame rotating at their average frequency and includes the gain and loss terms of \cref{eq:qvdp} for both oscillators.
For simplicity, we focus on oscillators with identical rates $\kappa$ and $\kappa_{1,2}$.
The frequency of each oscillator, however, can be different, which is parametrized by the detuning $\delta$.
Finally, the coupling is described by the dissipative interaction term $\diss[a-b]\rho$ with interaction strength $V$.

Deriving the mean-field equations, we find the same equations as those of the two coupled classical vdP oscillators, Eqs.~\eqref{eq:2vdp_alpha}, with $\alpha = \expval{a}$, $\beta = \expval{b}$.
We therefore expect phase locking and frequency entrainment as discussed for the classical case in \cref{s:sync_two_coupled}. 
These expectations will be confirmed in the following by analyzing the master equation~\eqref{eq:2qvdp}.

\begin{figure*}
    \centering
    \includegraphics[width=2.8in]{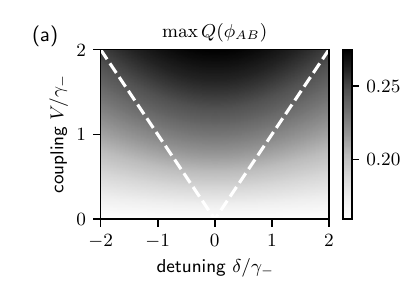}%
    \includegraphics[width=3in]{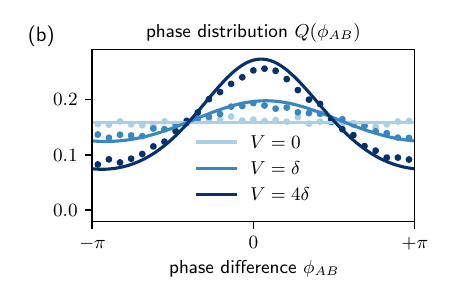}
    \caption{Phase locking of quantum vdP oscillators.
    (a)
    The grayscale shows the maximum of the phase distribution $\max Q(\phi_{AB})$, a measure of phase locking.
    The white dashed line indicates the synchronization transition $V=\abs{\delta}$ expected in classical noiseless oscillators.
    (b)
    The lines show the phase distribution $ Q(\phi_{AB})$ for different values of the coupling strength.
    The scatter points show the corresponding estimate of the phase distribution obtained via the heterodyne current; see \cref{eq:phiab_meas}.
    For each value of the coupling strength, they are computed from one trajectory of duration $5000/\delta$ with 40\,000 uniformly spaced samples.
    Parameters: $\kappa=\kappa_2$, $\kappa_1 = 3\kappa_2$. (b) $\delta = \kappa / 2$.
    }
    \label{fig:2qvdp_measured_phase}
\end{figure*}

\subsubsection{Phase locking}
Various measures for phase locking of quantum oscillators have been suggested~\citep{ludwig_marquardt2013,lee_QuantumSynchronizationQuantum_2013,walterQuantumSynchronizationDriven2014,Fazio2013,Hush_2015,Weiss_2016,Jaseem2020}.
Here, we use the Husimi-Q distribution, extending the definition of \cref{eq:qvdP_Q} to two oscillators:
\begin{equation}
    Q(\alpha, \beta) = \frac{1}{\pi}\bra{\alpha} \otimes \bra{\beta} \rho \ket{ \alpha} \otimes \ket{\beta } \, ,
\end{equation}
projecting the density operator on the coherent states $\ket{\alpha}$ and $\ket{\beta}$.
From $Q(\alpha, \beta)$, we derive the distribution $Q(\phi_{AB})$ of the phase difference $\phi_{AB} = \phi_A - \phi_B$.
It is obtained by integrating out the radial degree of freedom as well as the total phase,
\begin{align}
    Q(&\phiab) = 
    \int_0^{2\pi} \d \phi_A
    \d \phi_B
    \int_0^{\infty} \d r_A r_A \d r_B r_B
    \label{eq:2qvdP_Q_phasedistribution}
    \\
    \times & Q(\alpha = r_A e^{-i\phi_A}, \beta = r_B e^{-i\phi_B})
    \delta(\phiab - \phi_A + \phi_B)
    \, .
    \nonumber
\end{align}

\Cref{fig:2qvdp_measured_phase}(b) displays the phase distribution by the solid lines for different coupling strengths.
In the absence of coupling, the phase distribution $Q(\phi_{AB})$ is completely flat, i.e., all phase differences are equally likely, indicating the absence of phase locking.
When increasing the coupling strength, $Q(\phi_{AB})$ develops an increasingly large peak, indicating partial phase locking; it becomes more likely for the phase difference to take a value close to zero.
To quantify the degree of synchronization by a single number, we use the maximum value of the phase distribution, $\max Q(\phi_{AB})$.
It is shown in \cref{fig:2qvdp_measured_phase}(a) as a function of coupling strength and detuning.
The amount of phase locking increases with $V/\abs{\delta}$.
The white dashed lines indicate the synchronization threshold $V=\abs{\delta}$ for the classical analog without fluctuations.
The phase-locking behavior of two coupled quantum vdP oscillators is qualitatively the same as that of two classical oscillators in the presence of fluctuations; compare to \cref{fig:classical_phase_locking}.

\subsubsection{Frequency entrainment}
\begin{figure}
    \centering
    \includegraphics[width=3in]{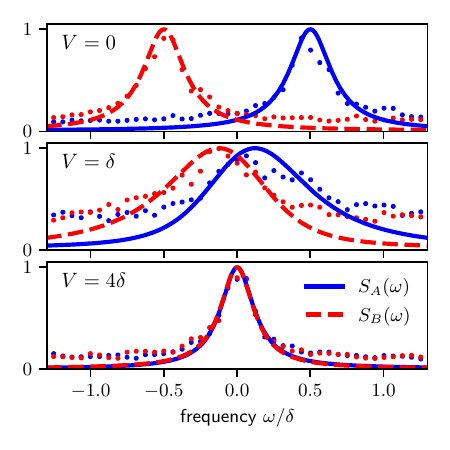}
    \caption{
    Frequency entrainment of quantum vdP oscillators.
    The lines show the spectra as calculated by \cref{eq:2qvdp_spectra} for oscillators $A$ (blue solid) and $B$ (red dashed).
    The scatter points show the spectra calculated from the heterodyne currents; see \cref{eq:2qvdp_spectra_meas}. 
    The heterodyne spectra are averaged over a window of width $\delta / 10$.
    Parameters: $\kappa=\kappa_2$, $\kappa_1 = 3\kappa_2$, $\delta = 5 \kappa_2$.
    Each spectrum is computed from one trajectory of duration $5000/\delta$ with 40\,000 uniformly spaced samples.
    }
    \label{fig:2qvdp_measured_frequency}
\end{figure}

Frequency entrainment of the two oscillators can be analyzed via their spectra~\citep{Walter_2015}.
The spectra of classical oscillators were introduced in \cref{s:sync_two_coupled}; see \cref{eq:spectra_classical}.
For quantum oscillators, the steady-state spectra are defined as \citep{Breuer} 
\begin{equation}
\begin{split}
    S_A(\omega) & = \lim_{t\rightarrow \infty} \int_{-\infty}^{\infty} \d \tau \expval*{a^\dag(t+\tau) a(t)}e^{-i\omega \tau}
\end{split}
\label{eq:2qvdp_spectra}
\end{equation}
(and analogously for oscillator $B$),
i.e., the Fourier transforms of the two-time correlations $\expval*{a^\dag(t+\tau) a(t)}$ and $\expval*{b^\dag(t+\tau) b(t)}$.
The spectra of the two coupled quantum vdP oscillators are shown by the lines in \cref{fig:2qvdp_measured_frequency}.
We find a behavior similar to that of classical vdP oscillators; compare to \cref{fig:classical_frequency_entrainment}(b):
Each spectrum features a peak whose width stems from the phase diffusion and whose position represents the typical frequency.
The frequency difference is $\delta$ in the absence of a coupling and approaches zero when increasing the coupling.

In summary, we have shown that the synchronization behavior of two coupled quantum vdP oscillators is qualitatively the same as that of two classical oscillators in the presence of fluctuations.

\subsubsection{Quantum synchronization under heterodyne detection}
We now show that both phase locking and frequency entrainment can be observed via heterodyne detection.
Consider the evolution of the two quantum vdP oscillators in \cref{eq:2qvdp} with an additional independent measurement for each oscillator,
\begin{align}
    \label{eq:2qvdp_meas}
    \dt \rho_\m =
    &-i\frac{\delta}{2}[ a^\dag a - b^\dag b, \rho_\m ]
    + \kappa_1 (\diss[a^\dag] +\diss[b^\dag])\rho_\m
    \notag \\
    & +
    \kappa_2 (\diss[a^2]+\diss[b^2])\rho_\m
    + V \diss[a - b]\rho_\m 
    \notag \\
    &
    +\kappa (\diss[a]+\diss[b]) \rho_\m 
    \\
    & + 
    \frac{\mathrm{d}W_A}{\d t}
    \sqrt{\kappa} 
    \left[
    e^{i\omega_\m t}(a - \Tr[a\rho_\m]) \rho_\m + \mathrm{H.c.}
    \right]
    \notag \\
    & + 
    \frac{\mathrm{d}W_B}{\d t}
    \sqrt{\kappa} 
    \left[
    e^{i\omega_\m t}(b - \Tr[b\rho_\m]) \rho_\m + \mathrm{H.c.}
    \right]
    \, .
    \notag
\end{align}
In an experiment, one observes the heterodyne current~\citep{Wiseman_2009}
\begin{equation}
\begin{split}
    I_\mathrm{het,A} &= \sqrt{\kappa}\Tr[a \rho_\m] + {\frac{1}{\sqrt{2}}} \left(\frac{\d W_{A,x}}{\d t} + i \frac{\d W_{A,y}}{\d t} \right) \, 
\end{split}
\end{equation}
(and analogously for oscillator $B$).
The current is determined by the expectation values of $a$ and the detector noise, which is described by the independent Wiener increments $\d W_{A,x}$ and $\d W_{A,y}$ that induce fluctuations in the real and imaginary part of the amplitude.

The currents can be used to approximate the measures of phase locking and frequency entrainment, i.e., the phase distribution $Q(\phiab)$ and the spectra $S_{A,B}(\omega)$.
The phase difference between the two oscillators is estimated by the phase relation of the currents
\begin{equation}
    \phiab^\m = \arg[I_\mathrm{het,B} / I_\mathrm{het,A}]
    \, .
    \label{eq:phiab_meas}
\end{equation}
The phase distribution can be approximated by the normalized frequency of occurrence of $\phiab^\m$ in the stationary state, which is shown in \cref{fig:2qvdp_measured_phase}(b) by the scatter points.
While an exact agreement between the estimated phase distribution and $Q(\phiab)$ is not expected, they qualitatively agree very well.

The heterodyne currents can moreover be used to estimate the spectra via
\citep[Section 4.5.1]{Wiseman_2009}
\begin{equation}
\begin{split}
    S^\m_A(\omega) &= \lim_{t\rightarrow\infty}
    \int_{-\infty}^\infty 
    \d \tau e^{-i\omega \tau}
    \mathbb{E}[I_\mathrm{het,A}(t+\tau)^* I_\mathrm{het,A}(t)]
    \\
    &=
    \kappa S_A(\omega) + 1
    \label{eq:2qvdp_spectra_meas}
    \end{split}
\end{equation}
and analogously for oscillator $B$.
The constant term stems from the white noise of the detector partially masking the signal.
The measured spectra are shown in \cref{fig:2qvdp_measured_frequency} by the scatter points.
They approximate well the spectra $S_{A,B}(\omega)$ displayed by the lines despite the presence of the white noise floor.

In summary, we showed that heterodyne detection presents a tool to approximately measure both phase locking and frequency entrainment of coupled quantum limit cycles.

\begin{figure*}
    \centering
    \includegraphics[width=5.8in]{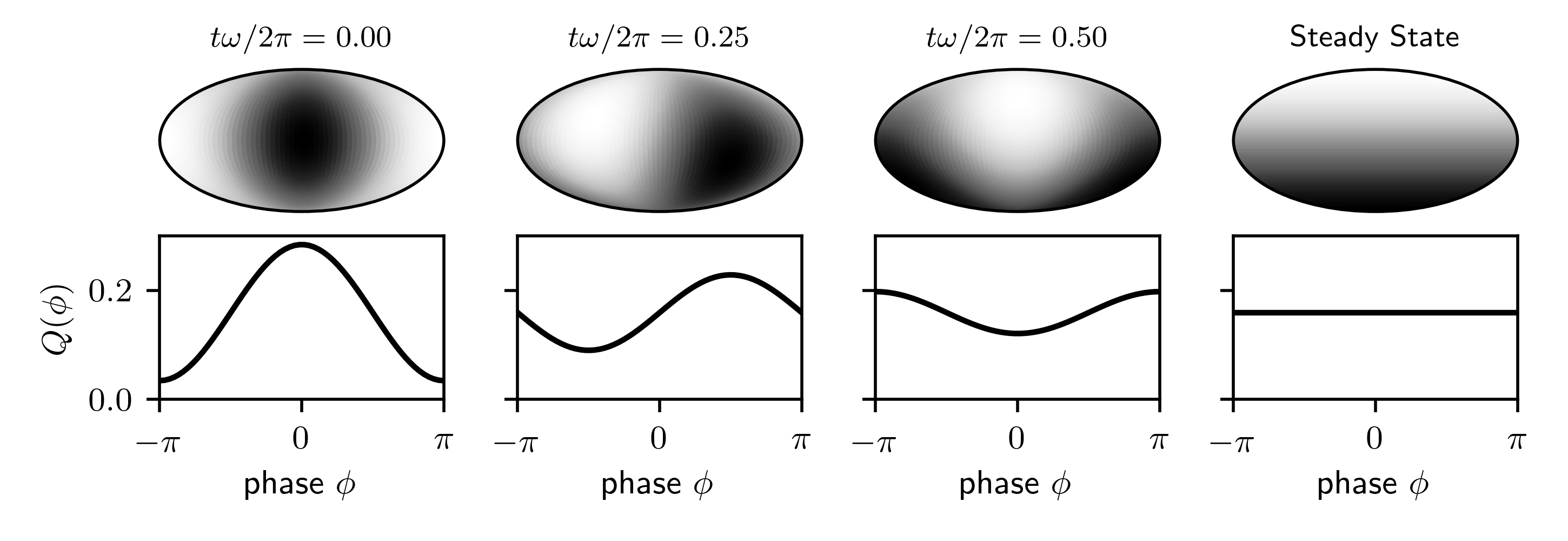}
    \caption{Time evolution of the spin-$1/2$ oscillator shown by the Husimi-Q distribution $Q(\theta,\phi)$ (top row) and the phase distribution $Q(\phi)$ (bottom row).
    The Husimi-Q distribution is shown by a Mollweide projection: Lines of constant $\theta$ are horizontal; the top and bottom of the plot correspond to $\theta = 0$ and $\theta = \pi$, and $\phi$ increases from left to right.
    Parameters: $\gp = \gm/2, \omega = 2\gm$.}
    \label{fig:spin_LC}
\end{figure*}

\subsection{Discussion}

A main contribution of this paper is highlighting the conceptual similarity between classical and quantum limit cycles as well as their synchronization properties through the measurement perspective, as we have shown in Sections~\ref{s:qvdp_qlc} and \ref{s:qvdp_2}.
The quantum trajectories obtained under heterodyne detection make the limit-cycle dynamics directly visible, displaying the same qualitative behavior as classical limit cycles subject to noise: persistent oscillations with stabilized amplitude.

At the ensemble level, the density operator governed by the master equation plays an analogous role as the probability distribution of the classical limit cycle subject to noise governed by the Fokker-Planck equation, Eq.~\eqref{eq:FP_vdP}.
In phase space, both exhibit a static, ring-shaped steady state.
This parallel is important in understanding what constitutes a quantum limit cycle.
Rather than considering only the ensemble average, our work suggests examining individual trajectories.
More specifically, we suggest examining trajectories obtained under heterodyne measurement, since this type of measurement provides information about both quadratures and is thus a natural choice in characterizing limit cycles.

While we focused on the similarity between classical and quantum limit cycles, we expect that the measurement perspective can reveal new insights into quantum limit cycles without a classical analog, see for example Refs.~\cite{Lorch_2014,rouletSynchronizingSmallestPossible2018}.
Moreover, in coupled limit-cycle oscillators, there are genuine quantum synchronization effects, for example synchronization blockades~\citep{lorchQuantumSynchronizationBlockade2017,lorchGenuineQuantumSignatures2016} or a boost of synchronization via dissipation~\citep{Mok_2020}.
Extending our quantum trajectory analysis to these systems presents interesting possibilities for future studies.

Beyond  conceptual insights into quantum limit cycles, we showed that heterodyne detection is a tool to approximately measure both phase locking and frequency entrainment.
This connects previously suggested theoretical measures of quantum synchronization with quantities accessible in experimental settings.
Previous experiments typically used quantum state tomography or Wigner function reconstruction to observe synchronization and focused on ensemble averages~\cite{koppenhoferQuantumSynchronizationIBM2020,Liu_2025,Li_2025,Behrle_2023,zhang2023observing,PhysRevA.105.062206,PhysRevLett.125.013601}.
Continuous measurement using heterodyne detection provides a different approach, which additionally allows the observation of the trajectory dynamics of limit-cycle oscillators.
Heterodyne detection has already been successfully employed in quantifying synchronization of atomic ensembles~\citep{Cox_2014,Weiner_2017,Natale_2025}.

\section{Quantum spins}
\label{s:quspins}

We now turn our attention to two-level systems, i.e., spins-$1/2$, which only comprise two states $\ket{0}$ and $\ket{1}$.
It has been debated whether such systems can host limit cycles~\citep{rouletSynchronizingSmallestPossible2018,parra-lopezSynchronizationTwolevelQuantum2020,zhang2023observing}.
We will conduct a similar analysis as in the previous section and use the measurement perspective to address the debate of limit cycles in spin-$1/2$ systems.

\subsection{Quantum limit cycles}
The only gain and loss processes that are possible in a spin-$1/2$ system are linear; a nonlinear two-excitation loss that is key in the dynamics of a quantum vdP oscillator cannot exist in a two-level system.
We will therefore consider the master equation
\begin{equation}
    \dot \rho
    =
    -i[\frac{\omega}{2} \sigma^z,\rho] +
    \gamma_+ \mathcal{D}[\sigma^+]\rho  + 
    \gamma_- \mathcal{D}[\sigma^-]\rho 
	\label{eq:sync_spin_LC}
\end{equation}
with frequency $\omega$, gain rate $\gamma_+$, and loss rate $\gamma_-$.
Here, and in the following, we use Pauli matrices $\sigma^{x,y,z}$ and ladder operators $\sigma^+=\dyad{1}{0}$, $\sigma^- = \dyad{0}{1}$ (when there are two spins, we add subscripts $A,B$).
A spin-$1/2$ has three degrees of freedom $s^{\pm,z} = \expval{\sigma^{\pm,z}}$, and the master equation~\eqref{eq:sync_spin_LC} can be rewritten as
\begin{align}
    \dt s^z &= (\gamma_+ - \gamma_-) - (\gamma_+ + \gamma_-) s^z 
    \, ,
    \\ 
    \dt s^+ &= i\omega s^+ - (\gamma_+ + \gamma_-) s^+/2 
    \, .
\end{align}
The first line describes the dynamics of the population $s^z$, which approaches $(\gamma_+ - \gamma_-)/(\gamma_+ + \gamma_-)$ in the steady state.
The second line describes the oscillation at frequency $\omega$ and the decoherence at rate $(\gp+\gm)/2$.
As a result of decoherence, $s^+$ vanishes in the long-time limit.
\begin{figure*}
    \centering
    \includegraphics[width=2in]{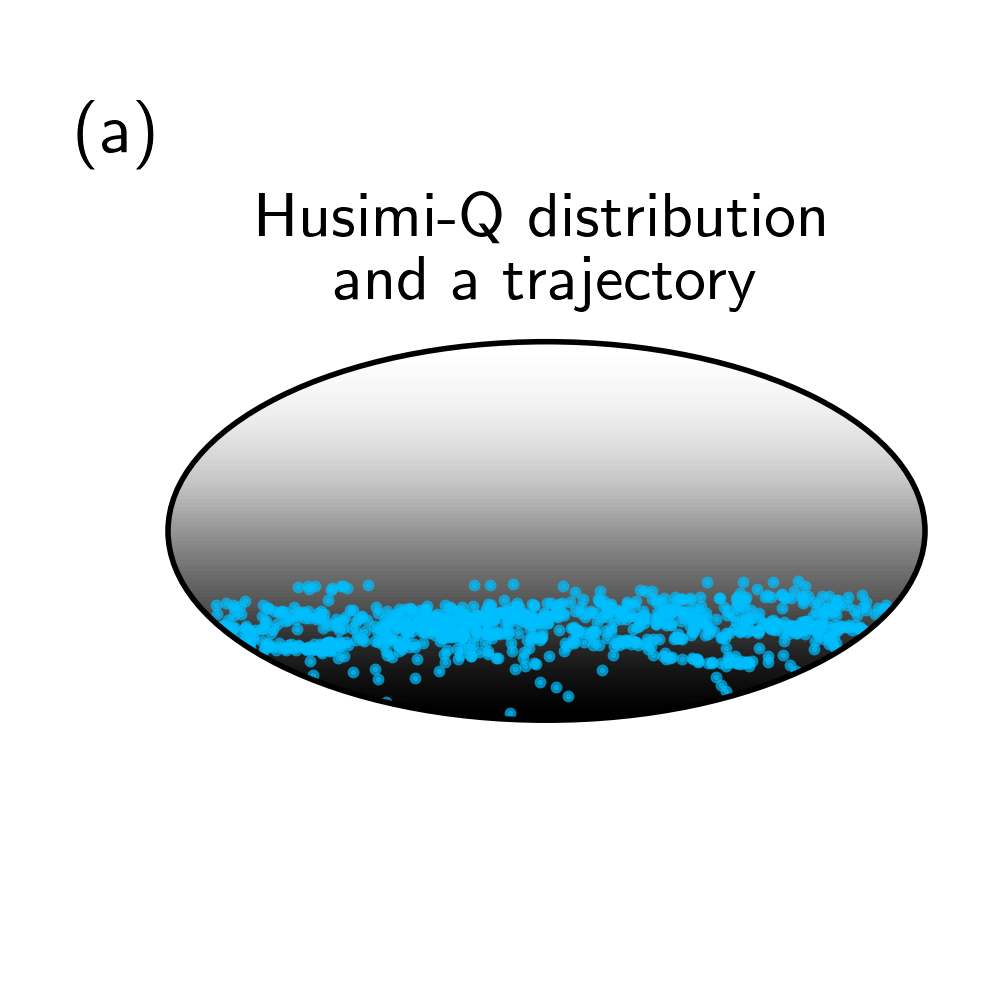}%
    \includegraphics[width=3in]{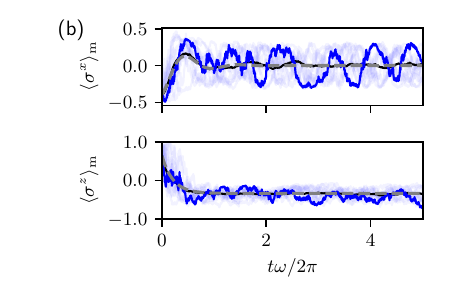}
    \caption{
    Spin-1/2 oscillator: Limit cycle under measurement.
    (a)~%
    The heatmap shows the steady-state Husimi-Q distribution from \cref{fig:spin_LC}, top-right panel.
    The blue scatter points show the time evolution of one quantum trajectory.
    (b)
    The light blue lines show the expectation values $\expval{\sigma^x}_\m$ and $\expval{\sigma^z}_\m$ of ten different quantum trajectories.
    The dark blue line highlights one of them.
    The black line shows the average of $100$ such realizations.
    It overlaps well with the gray dashed line, which shows $\expval{\sigma^x}$ obtained from the solution of the master equation without measurement.
    Same parameters as in \cref{fig:spin_LC}: $\gp = \gm/2, \omega = 2\gm$.
    }
    \label{fig:spin_LC_measurement2}
\end{figure*}

In terms of the density operator, the steady state of the system is
\begin{equation}
    \rho_\mathrm{ss} = \frac{1}{\gamma_+ + \gamma_-}
    \left( \gamma_- \dyad{0} + \gamma_+ \dyad{1} \right) 
    \, .
    \label{eq:sync_spin_LC_ss}
\end{equation}
This state is a mixture of the states $\ket{0}$ and $\ket{1}$ and the possibility of synchronization in two-level systems was dismissed in \cite{rouletSynchronizingSmallestPossible2018}.
However, it was shown that two-level systems can 
exhibit features of synchronization~\cite{parra-lopezSynchronizationTwolevelQuantum2020}, which have been observed in a trapped ion subject to an external drive~\citep{zhang2023observing}.
Our analysis will support the idea that 
in the presence of heterodyne detection, spins-$1/2$ can be considered as quantum limit-cycle oscillators.

First, we discuss the phase distribution of a spin-$1/2$, which is analogous to the phase distribution of the classical and quantum vdP oscillators.
It is defined using spin-coherent states~\citep{rouletSynchronizingSmallestPossible2018, parra-lopezSynchronizationTwolevelQuantum2020}
\begin{equation}
    \ket{\theta,\phi} =
    \exp(-i\phi \sigma^z/2)
    \exp(-i\theta \sigma^y/2) \ket{1} \, ,
\end{equation}
and the Husimi-Q function 
\begin{equation}
    Q(\theta,\phi) = \frac{1}{2\pi} \expval{\rho}{\theta,\phi} \, ,
    \label{eq:spin_Husimi_Q}
\end{equation}
where $\phi$ is the azimuthal phase, which is the relevant quantity in synchronization.
The parameter $\theta$ determines the population difference:
$\expval{\sigma^z}{\theta,\phi}=
\cos (\theta)$.
The distribution of the phase $\phi$ is obtained by integrating over $\theta$,
\begin{equation}
    Q(\phi) = \int_0^\pi \mathrm{d} \theta \sin \theta Q(\theta,\phi)
    = \frac{1}{2\pi} + \frac{1}{4} \Re[\expval{\sigma^+}e^{-i\phi}]
    \, .
    \label{eq:spin_Husimi_Q_phase}
\end{equation}

The dynamics of the master equation \eqref{eq:sync_spin_LC} are visualized in \cref{fig:spin_LC}.
Similar to the analyses of classical limit-cycle oscillators in the presence of fluctuations [see \cref{fig:vdP_noise}] and the quantum vdP oscillator [see \cref{fig:qvdp_time_evolution}], we identify three processes taking place:
the rotation at frequency $\omega$,
the attraction to the limit cycle,
and the phase diffusion.

\subsubsection{Quantum limit cycles under heterodyne detection}

As for the quantum vdP oscillator, the spin's limit-cycle structure becomes apparent through heterodyne detection.
In the presence of a measurement of $\sigma^-$, the master equation is
\begin{align}
    \dot \rho_\m
    =&
    -i[\frac{\omega}{2} \sigma^z,\rho_\m] +
    \gamma_+ \diss[\sigma^+] \rho_\m +
    \gamma_- \diss[\sigma^-] \rho_\m +
	\label{eq:sync_spin_LC_measurement}
    \\
    &+\frac{\mathrm{d}W}{\d t}
    \sqrt{\gamma_-} 
    \left[
    e^{i\omega_\m t}(\sigma^- - \Tr[\sigma^-\rho_\m]) \rho_\m + \mathrm{H.c.}
    \right]
    \, ,
    \nonumber
\end{align}
with the usual statistics of the noise process $\d W$; see \cref{eq:1qvdp_meas} and the paragraph below.
Integrating this master equation, we obtain quantum trajectories for the spin-$1/2$ oscillator;
several of them are shown in \cref{fig:spin_LC_measurement2}.
Despite the presence of fluctuations, we see that the expectation value $\expval{\sigma^z}_\m$ approaches a constant value, while the expectation value $\expval{\sigma^x}_\m$ displays the key characteristics of limit cycles: persistent oscillations with stabilized amplitude.
Therefore, we conclude that the spin-$1/2$ system under heterodyne measurement shows limit-cycle oscillations.

On the other hand, the ensemble average (black line) displays a damped oscillation that agrees well with the solution of the master equation (gray dashed line).
This behavior is equivalent to both the quantum vdP oscillator [see \cref{fig:qvdp_LC_measurement}] and a classical noisy limit-cycle oscillator [see \cref{fig:vdp_classical_trajectories}].

The identification of limit-cycle trajectories under measurement is a remarkable result, given that the master equation describes a thermal state when $\gamma_- > \gamma_+$.
We will further discuss the presence of limit cycles in quantum trajectories and in the master equation of a spin-$1/2$ in \cref{s:discussion_spins}.

\begin{figure*}
    \centering
    \includegraphics[width=2.8in]{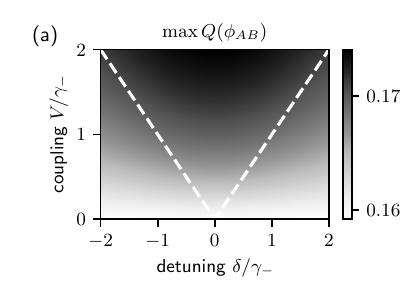}%
    \includegraphics[width=3in]{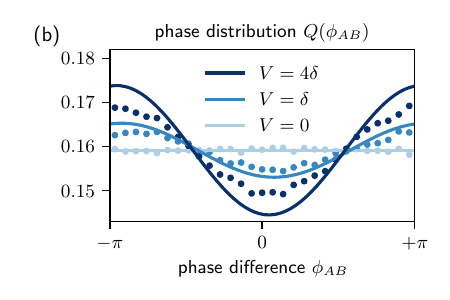}
    \caption{
    Phase locking of spin-$1/2$ oscillators.
    (a)
    The grayscale shows the maximum of the phase distribution $\max Q(\phi_{AB})$, a measure of phase locking.
    The white dashed line indicates the synchronization transition $V=\abs{\delta}$ expected in classical noiseless oscillators.
    (b)
    The lines show the phase distribution $Q(\phi_{AB})$ for different values of the coupling strength.
    The scatter points show the corresponding estimate of the phase distribution obtained via the heterodyne current; see \cref{eq:phiab_meas_spins}.
    For each value of the coupling strength, they are computed from one trajectory of duration $5\times 10^5/\delta$ with $4\times 10^6$ uniformly spaced samples.
    While the lines and the scatter points do not perfectly overlap, they show a qualitative agreement.
    Parameters: $\gp=\gm/2$. (b) $\delta = \gm/2$. 
    }
    \label{fig:2spins_phase}
\end{figure*}

\subsection{Two coupled spins}
Let us now discuss synchronization of two coupled spin-$1/2$ oscillators.
Their time evolution is governed by 
\begin{align}
    \dot \rho
    =
    &
    -i\frac{\delta}{4} [\sigma_A^z - \sigma_B^z,\rho]
    +
    V \mathcal{D}[\sigma^-_A + \sigma^-_B]\rho
	\label{eq:sync_two_spins}
    \\
    \nonumber
    &+\gamma_+ \left( \mathcal{D}[\sigma^+_{A}]+\mathcal{D}[\sigma^+_{B}]\right)\rho 
    +\gamma_- \left( \mathcal{D}[\sigma^-_{A}]+\mathcal{D}[\sigma^-_{B}]\right)\rho 
    \, .
\end{align}
The system is described in the rotating frame of the average frequency.
The detuning between the two spins is $\delta$, and the coupling strength is $V$.
We can obtain intuition about the dynamics by considering the mean-field equations for the expectation values $s^{+,z}_{A,B} = \expval*{\sigma^{+,z}_{A,B}}$,
\begin{align}
    &\dt s^z_{A,B} = (\gamma_+ - \gamma_- - V) - (\gamma_+ + \gamma_- + V) s^z_{A,B} 
    \, ,
    \\ 
    &\dt s^+_A \approx [+i\delta s^+_A - (\gamma_+ + \gamma_- + V) s^+_A + Vs^z_A s^+_B]/2
    \, ,
    \\ 
    &\dt s^+_B \approx [-i\delta s^+_B - (\gamma_+ + \gamma_- + V) s^+_B + Vs^z_B s^+_A]/2
    \, .
\end{align}
We have approximated $\expval{\sigma^z_A \sigma^+_B} \approx s^z_A s^+_B$ and the same for $A\leftrightarrow B$. 
Within this approximation,
the phases $\phi_{A,B} = \arg[s^+_{A,B}]$ evolve as follows,
\begin{align}
    \dt \phi_{AB} = 
    \delta - \frac{V}{2} \left(
    s^z_B\frac{\abs{s^+_A}}{\abs{s^+_B}} + 
    s^z_A\frac{\abs{s^+_B}}{\abs{s^+_A}}
    \right)
    \sin(\phi_{AB}) 
    \, .
\end{align}
The phase interactions are qualitatively the same as the phase interactions between two classical limit-cycle oscillators [see \cref{eq:adler}] with an effective coupling strength that depends on the $s^z_{A,B}$ as well as $\abs*{s^+_{A,B}}$.
Therefore, we expect that a positive steady-state value of $s^z_{A,B}$ results in in-phase locking, while negative $s^z_{A,B}$ results in out-of-phase locking with phase difference $\pi$.
To confirm this expectation, we now solve the full master equation~\eqref{eq:sync_two_spins} to analyze phase locking.

\subsubsection{Phase locking}
We use a distribution for the phase difference similar to the analyses of the classical and quantum vdP oscillators.
It is obtained in analogy to \cref{eq:2qvdP_Q_phasedistribution} by first projecting the density matrix on spin-coherent states
\begin{equation}
\begin{split}
    &Q(\theta_A,\phi_A,\theta_B,\phi_B)
    = 
    \\
    &\frac{1}{4\pi^2}
    \bra{\theta_A,\phi_A} \otimes \bra{\theta_B,\phi_B}
    \rho
    \ket{\theta_A,\phi_A} \otimes \ket{\theta_B,\phi_B}
    \, .
\end{split}
\end{equation}
Then, we integrate over the polar angles $\theta_{A,B}$ and over the global phase to obtain a distribution of the phase difference $\phi_{AB} = \phi_A - \phi_B$
\begin{align}
    Q(\phi_{AB}) =&
    \int_0^\pi \mathrm{d} \theta_A \sin \theta_A
    \mathrm{d} \theta_B \sin \theta_B
    \int_0^{2\pi} \d \phi_A
    \d \phi_B
    \nonumber
    \\
    \label{eq:2spins_Qphase_diff}
    & \times Q(\theta_A,\phi_A,\theta_B,\phi_B)
    \delta(\phi_{AB} - \phi_A + \phi_B) \\
    =&\frac{1}{2\pi} + \frac{\pi}{16}\Re[\expval{\sigma^+_A \sigma^-_B}e^{i\phiab}]
    \, .
    \nonumber
\end{align}

The maximum value of $Q(\phi_{AB})$ and the phase distribution are shown in \cref{fig:2spins_phase}.
Qualitatively, we find the same behavior as for the quantum vdP oscillator and the classical noisy oscillator; see \cref{fig:2qvdp_measured_phase} and \cref{fig:classical_phase_locking}.
Here, the spins phase-lock with a phase difference of $\pi$, since we chose parameters where the gain is smaller than the loss, so that $s^z$ assumes negative values.
Quantitatively, the synchronization measure is in general smaller compared to that of quantum vdP oscillators due to the smaller Hilbert space.
For the spins, the maximum deviation of the phase distribution from its average $1/(2\pi)$ is $\pi/32$ where $\expval{\sigma^+_A\sigma^-_B} = 1/2$.
For the quantum and classical vdP oscillators, it is not bounded.

\subsubsection{Frequency entrainment}
\begin{figure}
    \centering
    \includegraphics[width=3in]{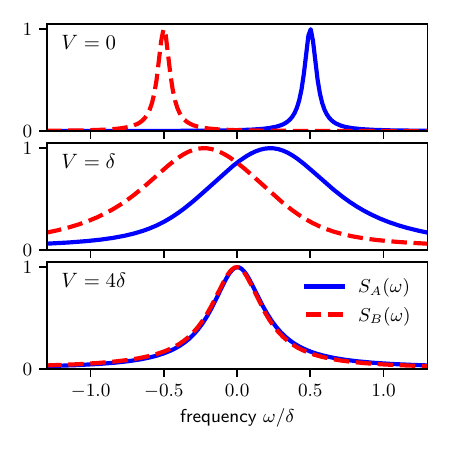}%
    \caption{
    Frequency entrainment of spin-$1/2$ oscillators.
    The lines show the spectra as calculated by \cref{eq:2spins_spectra} for oscillators $A$ (blue solid) and $B$ (red dashed) normalized to their maximum value.
    Parameters: $\gp=\gm/2$, $\delta = 5\gm$. 
    }
    \label{fig:2spins_frequency_locking}
\end{figure}

To analyze frequency entrainment of two detuned spins, we calculate the spectra in analogy to \cref{eq:2qvdp_spectra},
\begin{equation}
\begin{split}
    S_{A,B}(\omega) & = \lim_{t\rightarrow \infty} \int_{-\infty}^{\infty} \d \tau \expval*{\sigma^+_{A,B}(t+\tau) \sigma^-_{A,B}(t)}e^{-i\omega \tau}
\end{split}
\label{eq:2spins_spectra}
\end{equation}
i.e., the Fourier transforms of the two-time correlations $\expval*{\sigma^+_{A,B}(t+\tau) \sigma^-_{A,B}(t)}$.
The spectra are shown in \cref{fig:2spins_frequency_locking} and display frequency entrainment in the same way as the quantum and noisy classical vdP oscillators; see \cref{fig:2qvdp_measured_frequency} and \cref{fig:classical_frequency_entrainment}. 

\subsubsection{Quantum synchronization under heterodyne detection}
The analysis of synchronization of two spins under heterodyne detection is carried out in the same way as for the two quantum vdP oscillators.
The conditional master equation is
\begin{align}
    \dot \rho_\m
    =
    &-i\frac{\delta}{4} [\sigma_A^z - \sigma_B^z,\rho_\m]
    +
    V \mathcal{D}[\sigma^-_A + \sigma^-_B]\rho_\m
    \nonumber \\
    &+\gamma_+ \left( \mathcal{D}[\sigma^+_{A}]+\mathcal{D}[\sigma^+_{B}]\right)\rho_\m 
    +\gamma_- \left( \mathcal{D}[\sigma^-_{A}]+\mathcal{D}[\sigma^-_{B}]\right)\rho_\m
    \nonumber \\
    &+\frac{\mathrm{d}W_A}{\d t}
    \sqrt{\gamma_-} 
    \left[
    e^{i\omega_\m t}(\sigma_A^- - \Tr[\sigma^-_A\rho_\m]) \rho_\m + \mathrm{H.c.}
    \right]
    \nonumber \\
    &+\frac{\mathrm{d}W_B}{\d t}
    \sqrt{\gamma_-} 
    \left[
    e^{i\omega_\m t}(\sigma^-_B - \Tr[\sigma^-_B\rho_\m]) \rho_\m + \mathrm{H.c.}
    \right]
    \, .
	\label{eq:sync_two_spins_measurment}
\end{align}
The heterodyne currents,
\begin{equation}
\begin{split}
    I_\mathrm{het,A} &= \sqrt{\gamma_-}\Tr[\sigma_A^- \rho_\m] + \frac{1}{\sqrt{2}} \left(\frac{\d W_{A,x}}{\d t} + i \frac{\d W_{A,y}}{\d t} \right) 
\end{split}
\end{equation}
(and analogously for spin $B$),
can be used to measure phase locking.
The phase difference $\phiab$ between the two oscillators is estimated by the phase relation of the currents
\begin{equation}
    \phiab^\m = \arg[I_\mathrm{het,B} / I_\mathrm{het,A}]
    \, .
    \label{eq:phiab_meas_spins}
\end{equation}
The relative number of occurrences of the measured phase is shown by the scatter points in \cref{fig:2spins_phase}.
They qualitatively follow the phase distribution $Q(\phiab)$.
While in principle, the spectra can be reconstructed from the heterodyne currents, in practice, the signal-to-noise ratio is small (therefore, the reconstructed spectra are not displayed in \cref{fig:2spins_frequency_locking}).
The signal-to-noise ratio can be increased, however, by considering ensembles of two-level systems~\citep{Cox_2014,Weiner_2017}.

\subsection{Discussion}
\label{s:discussion_spins}

We showed that individual quantum trajectories of a spin-$1/2$ obtained under heterodyne detection display the characteristic features of limit cycles: persistent oscillations with stabilized amplitude.
Remarkably, under measurement, the spin can display a stabilized amplitude even with linear gain and loss processes.
Our analysis proves useful regarding the question whether or not quantum spins-$1/2$ can host limit cycles~\citep{rouletSynchronizingSmallestPossible2018,parra-lopezSynchronizationTwolevelQuantum2020,zhang2023observing}.
While these works mostly focused on the ensemble-averaged density operator, we have shown that continuous measurement using heterodyne detection offers a natural way to obtain quantum trajectories that make the limit cycle apparent.

One may further ask whether the master equation itself, \cref{eq:sync_spin_LC}, should be regarded as describing a limit-cycle system.
From the perspective of continuous measurement under heterodyne detection, one would answer the question in the affirmative, as the master equation describes the ensemble of all possible quantum trajectories and they show limit-cycle oscillations. 
However, under a different type of measurement, such as direct photodetection, the limit-cycle dynamics would not be visible.
The ambiguity exists because a quantum system's dynamics fundamentally depends on how the system is observed.
This point of view corresponds to considering quantum trajectories as ``subjectively real''~\citep{Wiseman_1996}, which
in the context of limit cycles implies that a master equation does or does not feature limit cycles depending on the measurement.

For experimental implementations, heterodyne detection can in principle be used to access the limit-cycle dynamics of spin systems.
While the observation of single spins may be challenging due to the large amount of noise, ensembles of spins have already been observed in this way~\citep{Cox_2014,Weiner_2017}.

\section{Conclusions}
\label{s:conclusions}

We have shown that individual quantum trajectories obtained through heterodyne detection exhibit persistent oscillations with stabilized amplitude analogous to the trajectories of classical limit-cycle oscillators subject to noise.
For both quantum van der Pol oscillators and spins-$1/2$, we analyzed limit cycles not only at the ensemble level but also on the trajectory level, similar to how classical limit cycles can be described in terms of a Fokker-Planck equation and the corresponding Langevin equation.

Viewing quantum limit cycles from a measurement perspective goes beyond the ensemble analysis based on the density operator.
Thus, our work sheds light onto the conceptual issue of what constitutes a quantum limit cycle.
While in our examples, heterodyne detection made the limit cycles apparent, other measurement strategies, such as direct photon counting, do not necessarily reveal the continuous limit-cycle dynamics.
Therefore, a system's master equation alone without specifying the coupling to the environment or a specific measurement type is not sufficient to decide whether it exhibits a limit cycle.
One may say that a master equation features a limit cycle if there is at least one measurement or unraveling that renders it apparent.

For coupled limit-cycle oscillators, we demonstrated that synchronization of both quantum and classical limit-cycle oscillators can be characterized by analogous measures of phase locking and frequency entrainment.
These measures are based on phase-space distributions and emission spectra.
While these measures have been discussed theoretically, this work shows their connection to an experimentally accessible quantity: the heterodyne current of the photodetector.
Whereas previous experiments on quantum synchronization have relied on quantum state tomography or Wigner function reconstruction,
heterodyne detection presents a different approach, which additionally provides insights into the dynamics of the limit cycles.

There are several interesting questions for future studies.
For one, we only focused on the case of $U(1)$-symmetric master equations.
One may also consider symmetry-breaking terms such as a coherent drive, which can induce synchronization.
The measurement perspective discussed will also apply to this scenario.
Moreover, while we focused on the quantum vdP oscillator and quantum spins-$1/2$, the measurement perspective should also prove useful for other types of quantum limit cycles.
Of particular interest are cases where quantum synchronization differs from classical synchronization.
Much research is devoted to this fascinating topic, and we expect that the continuous-measurement perspective provided here will facilitate the analysis of quantum effects in synchronization.

\begin{acknowledgments}
We thank Matteo Brunelli, Martin Koppenhöfer, and Niels Lörch for valuable discussions. 
In addition, we acknowledge financial support from a Swiss National Science
Foundation individual Grant (No.\ 200020 200481).
\end{acknowledgments}

\end{document}